\documentclass[floatfix, preprintnumbers, amsmath, amssymb, 10pt]{revtex4}
\usepackage{graphicx}
\usepackage{dcolumn}
\usepackage{bm}

\begin{document}

\date{\today}

\title{The Cosmology of Massless String Modes}

\author{Subodh P. Patil $^{1,2,3,4)}$} \email[email: ]{patil@het.brown.edu}
\author{Robert H. Brandenberger$^{2,1)}$} \email[email: ]{rhb@het.brown.edu}

\affiliation{1) Dept.of Physics, Brown University, 
Providence R.I. 02912, U.S.A.} 
\affiliation{2) Dept.of Physics, McGill University, Montr\'eal QC, H3A 2T8, 
Canada}
\affiliation{3) Theoretical Physics, Queen Mary University of London, London, 
E1 4NS, U.K.}
\affiliation{4) DAMTP, Cambridge University, Cambridge CB2 0WA, U.K.}

\begin{abstract}
We consider the spacetime dynamics of a gas of closed strings in the context 
of General Relativity in a background of arbitrary spatial dimensions. Our 
motivation is primarily late time String Gas Cosmology, where such a spacetime 
picture has to emerge after the dilaton has stabilized. We find that after 
accounting for the thermodynamics of a gas of strings, only string modes 
which are massless at the self-dual radius are relevant, and that they lead to
a dynamics which is qualitatively different from that induced by the modes 
usually considered in the literature. In the context of an ansatz with three
large spatial dimensions and an arbitrary number of small extra dimensions, we
obtain isotropic stabilization of these extra dimensions at the self-dual 
radius. This stabilization occurs for fixed dilaton, and is induced by
the special string states we focus on. The three large dimensions undergo a 
regular Friedmann-Robertson-Walker expansion. We also show that this 
framework for late-time cosmology is consistent with observational bounds. 
\end{abstract}

\maketitle

\newcommand{\eq}[2]{\begin{equation}\label{#1}{#2}\end{equation}}

\section{Introduction}

The String Gas Cosmology (SGC) program, initiated in \cite{BV,TV} (see also
\cite{Perlt}) is a string-motivated cosmological model within which it appears
possible to obtain a nonsingular cosmology in which a 
universe with three large spatial dimensions is dynamically generated 
from initial conditions in which all spatial dimensions have a similar
microphysical scale. The scenario is based on coupling a
gas of string and brane matter degrees of freedom to a dilaton-gravity
background geometry. As initial conditions, space is chosen to be
compact and of string length in all directions, and matter is taken to be
a hot gas with all string and brane degrees of freedom excited. The
specific stringy degrees of freedom which have both winding and
momentum quantum numbers play an important role in the dynamics of the
early universe. There has been a
steady stream of research on this scenario over recent years (see e.g.
\cite{Cleaver,Mairi,sa2,de3,de2,nyc4,sw3,timon,nyc3,sa1,kaya4,kaya3,ac3,de1,sw2,nyc2,tb,ac2,Watson2,sp,Watson3,sw1,Watson4,kaya2,kaya1,ac1,ab,nyc1,am,Edna,Biswas2,borunda,rador,mcinnes}). 

In spite of its above mentioned main successes, SGC  
has encountered important obstacles (some of them yet 
to be resolved) which stand in the way of SGC being a complete and 
testable model of the early universe. These obstacles arise in attempting to 
implement the ideas introduced in \cite{BV} and \cite{sa2} in realistic 
string/M-theory settings. Chief among them, was the observation that a 
dynamical dilaton proved crucial in stabilizing the extra dimensions
(radion stabilization was first considered in \cite{sw2}), and as 
such posed a problem for  stabilization in the present epoch (see e.g.
\cite{sw1,ab}). However, this 
was not so much a general conclusion rather than an observation about the 
particular role the dilaton played in the specific realizations studied. 
Several other outstanding issues concern the question of stability of the 
extra dimensions to fluctuatons (addressed in \cite{Watson2,Watson3} at
the level of linear cosmological perturbation theory), 
the phenomenology of having a space-filling 
fluid of strings to maintain stabilization (we do not want to overclose the 
universe, whilst still maintaining an effective stabilization mechanism), and 
the consistency in using a low-energy effective theory at compactification 
radii comparable to the string scale. Also, a concern has recently been
raised \cite{nyc1,am} concerning the ability of string winding interactions
to only annihilate in three large spatial dimensions.

It is the goal of this paper to report 
on recent work which not only hopes to resolve some of the outstanding 
difficulties faced by SGC, but also points to a resulting model of the 
universe which is surprisingly complete, and potentially testable by 
observation. As such, this report constitutes the first in a series of 
papers \cite{rhbsp,brown}, where this particular paper will primarily 
elaborate on a model of the universe where extra dimensions (not just
one extra dimension as in \cite{sp}) are stabilized at late times 
(i.e. assuming the dilaton is fixed) by a confining potential induced by
a string gas which contains modes which are massless at the self-dual
radius. A subsequent report will 
further investigate the resulting phenomenology and point to potentially 
testable predictions arising from the model introduced here 
\cite{rhbsp}. There are already preliminary indications that the model we 
are studying contains a stringy mechanism for the generation of 
metric fluctuations \cite{rhbsp} 
and may even offer applications to the construction of a nonsingular
realization \cite{brown} of the Ekpyrotic/cyclic  universe scenario 
\cite{st1,st2,st3}.   

\par

Key to the results of this paper (and the phenomenology which  
follows) is the special role played by massless string states, whose utility
has been generally neglected - see, however, the discussions in
\cite{sp,Watson4}. In the following introductory section we 
begin by clarifying the philosophy of SGC, and then give a preview of some of 
the special properties of the massless string states which explains their 
vital role in SGC. After discussing the outstanding 
problems of SGC, we offer our ideas for resolving these problems 
within our framework. In Section III, we set up our model, study the 
effects of a gas of strings in a background described by Einstein gravity 
(fixed dilaton), and derive the resulting spacetime dynamics. We show that it 
is possible to stabilize any number of extra dimensions, making crucial use of 
the massless string states. In Sections IV and V, we demonstrate that this 
stabilization mechanism leads to a phenomenologically acceptable late-time 
cosmology (modulo the outstanding issue of the horizon problem of the three 
large dimensions). In particular, we show that 
Friedmann-Robertson-Walker (FRW) expansion of the 
universe results after the extra dimensions have stabilized. 

\section{SGC: Philosophy, Progress and Problems}

String Gas Cosmology is a paradigm constructed along similar principles as the 
Standard Big Bang Cosmology (BBC). Beginning with the observation of spatial homogeneity
and isotropy of the universe on large scales, BBC is the cosmology that results when one 
couples a theory of space-time (Einstein gravity) to matter described as a set of ideal gas
fluids (or a homogeneous distribution of fields as in the Inflationary Universe scenario
\cite{Guth}, an extension of BBC which solves some but not all of its problems), 
and assumes ``hot" initial conditions, i.e. that all matter degrees of freedom
are highly excited. 

The philosophy of String Gas Cosmology is similar. Instead of a homogeneous set of
ideal gases or fields, one takes matter to be a gas of all string and brane degrees
of freedom which string theory admits, and the background space-time is
described not necessarily in terms of Einstein gravity, but in terms of the
particular gravity theory (e.g. dilaton gravity) which
depends on which corner of the string theory moduli space one picks to be in. 
Novel features of SGC compared to BBC include the existence of extra dimensions and
extra fields (form fields and scalars) and the existence of new symmetries (dualities) 
relating the various corners of moduli space. The existence of 
extended objects in the matter sector allows for a radically different coupling between 
matter and geometry as 
these objects feel the metric tensor in a manner different from what the naive
application of intuition from BBC would imply. The new matter degrees of freedom 
also generate their own new symmetries (e.g. target space (T) duality) and cause the 
dynamics of spacetime to be very 
different than in BBC. The peculiar features of string thermodynamics, such as the 
existence of a limiting Hagedorn temperature \cite{hag} also serve to ensure that any 
stringy cosmology implemented in the spirit of BBC will have qualitatively different 
features.

\par

What then are the consequences of attempting to do cosmology in the framework outlined 
above? In addition to the fact that T duality raises the hope to be able to
construct a nonsingular cosmology \cite{BV}, the
main result (and in fact the main motivation for the SGC program) is 
that it is possible to generate a 3+1 dimensional universe through the dynamical 
``decompactification" of an early universe where all directions are initially taken to be 
compact and of similar size \cite{BV,sa2,sw2}. The way this works can briefly be 
summarized as follows: suppose we begin with a 10-dimensional toroidal universe, where 
all dimensions start at or around the self dual radius ($R = \sqrt{\alpha'}$). 
Populate this universe with a gas of strings in thermal equilibrium. Due to the 
presence of closed strings wound around the various cycles of the 10-torus, there is an 
energy expenditure associated with any expansion 
of a particular cycle, given by the relation:
\begin{equation}
\label{e}
E^2 \sim w^2R^2 \, ,
\end{equation}    
where $R$ is the radius of the one-cycle and $w$ is the winding number of a string wound 
along that cycle. Hence, the expansion of all of the cycles of the 10-torus should be 
held in 
check by the presence of these wound strings, so long as there are enough of them around. 
However, we know that oppositely wound closed strings like to scatter into unwound strings 
through the tree level interaction $w + \bar{w} \to ...$ (where the ellipses denote 
unwound strings). If we begin near a state of thermal equilibrium, with the tree level 
string interactions maintaining a non-zero population of wound strings, we can then ask the 
question under what conditions is it possible to maintain this state of thermal equilibrium? 
A naive dimension counting argument \cite{BV} tells us that the scattering cross-sections 
that describe winding mode annihilation depend on 
the dimensionality of spacetime. For instance, on a two dimensional torus, wound strings 
are very likely to intersect at some point. The same is true in three dimensions. 
However, in more than three spatial dimensions, the subset in phase space of initial 
conditions for which two wound strings intersect is of measure zero:
strings will generically miss each other (for some recent concerns with this argument
see \cite{nyc1,am}). Hence, in a background of any number of spatial dimensions, one 
finds that in at most 3 spatial dimensions these strings can meet each other, and hence 
unwind. It is in these dimensions that the universe is then free to expand if there is an 
initial expansion in place, the interactions having frozen out in the other dimensions. 
This stringy explanation for the dimensionality of spacetime \cite{BV} 
has been generalised \cite{sa2} to a setting where in addition to strings, one has branes of 
various dimensions present in the initial state of the universe.            
\par

Implementing the above argument in realistic settings has been the focus of much of the 
literature on SGC to date. In addition to the successes of 
this program, there are many open issues which remain. Among the successes is the result 
that isotropy of the 3-dimensions which subsequently describe our universe is a consequence 
of the dynamics of SGC \cite{sw3}. Progress has also been made in implementing SGC 
for more realistic compactifications of string theory (which do not neccesarily admit 
topological 1-cycles-- a pre-requisite for the existence of winding modes) \cite{nyc4}. 
However, a major obstacle to the success of this program is the fact that in the 
realizations of the SGC scenario to date, a dynamical dilaton proves to be crucial in 
obtaining stabilization of the extra dimensions \cite{sw2}. An argument \cite{TV} for
the neccessity of a dynamical dilaton for the stabilization of the extra dimensions
is that in general relativity, matter tends to cause 
monotonic expansion or contraction of the universe:
\begin{equation}
\label{hubble}
H^2 \propto \rho \, ,
\end{equation} 
and hence obtaining stabilization ($H = 0$) would be problematic. 
However, this intuition is flawed since it is based on assuming that 
one has isotropy of all dimensions. If instead we assume isotropy in the 
non-compact and separate isotropy in the compact dimensions then the 
time-time Einstein equation becomes:
\begin{equation}
\label{00}
\rho = \frac{1}{16\pi G_D}\bigl[d(d-1)H^2 + p(p-1)\mathcal H^2 +2pd\mathcal H H\bigr] \, ,
\end{equation}
where $G_D$ is Newton's gravitational constant in the full space-time, 
$d$ is the number of spatial non-compact dimensions, $p$ the number of 
compact dimensions, $H$ the Hubble rate in the non-compact dimensions, and $\mathcal H$ the 
Hubble rate in the compact dimensions. Hence one could easily have an oscillating 
$\mathcal H$ provided $H$ is large enough. In the case of one extra dimension, this
was shown explicitly in \cite{sp}. 

Another reason why a dynamical dilaton was included in most of the previous work on SGC
is that in the Type II string theory corners of the M-theory moduli space the dilaton
appears at the same footing as the graviton. Hence, in the absence of a mechanism which
stabilizes the dilaton, this field should be taken to be dynamical. A dynamical dilaton
also is crucial in order for T-duality to be manifest. 

However, in the current universe there is no evidence for a dynamical dilaton, and hence,
if SGC is to make successful connection with current observations, it must be assumed
that the dilaton is stabilized (ideally, this should be a result of string cosmology
itself). A crucial question for SGC is, therefore, whether the extra dimensions remain
stable after the dilaton has been fixed. 

The main goal of this paper is to show that any number of extra dimensions can be 
stabilized in the context of General Relativity (GR) 
(i.e. without a dynamical dilaton), and hence to extend the 
results of SGC to be applicable to late time cosmology. We 
find preliminary indications that the resulting cosmology  
has rich phenomenological implications (this will be discussed in more depth in a followup 
report \cite{rhbsp}). In addition to being stable to radion fluctuations
there is a potential 
non-inflationary mechanism for generating metric fluctuations \cite{brown}.  

The workhorse of this cosmology are string modes which are massless at the
self-dual radius, and whose role was discussed in a previous paper where the
stabilization of a single extra dimension was studied \cite{sp}. These
modes are related to enhanced symmetries at the self-dual radius (see e.g.
\cite{Watson4,stanford} for more general discussions). We argue that these states
must play a crucial role in SGC, and neglecting them will lead to incorrect
conclusions about the cosmological evolution.

The role of these special states is also important in addressing a common objection to 
SGC (and all other approaches to string cosmology), namely questioning the 
use a low energy effective theory description (dilaton gravity or GR) in a situation when 
characteristic lengths are around the string scale. 
The loop expansion breaks down once one reaches curvature regimes 
comparable to the string scale, and hence calls into question the consistency of the 
effective theory we are interested in. However, we will 
be studying the physics of certain massless modes which appear at enhanced symmetry points 
(i.e. the self dual radius), and as such describe new degrees of freedom that appear at 
these special points. Since the existence of these new degrees of freedom will transcend 
higher order corrections, we feel justified in using their perturbative properties. Perhaps 
an equivalent QFT intuition would be that just as the photon mass remains zero due to gauge 
invariance after higher order loop corrections are taken into account, we expect the 
properties of these states to similarly transcend higher order loop corrections due to the 
enhanced symmetries associated with them.     

An open issue for SGC\cite{RE} is the concern
that the states which keep the radii of the extra dimensions confined might overclose the 
universe. This concern would indeed be valid if these states were 
Planck mass objects (like string states with only winding number). However, if 
stabilization is achieved by a fluid of strings which are massless at the self-dual
radius, and which from the point of view of the large spatial dimensions behave as a
gas of massless states, the overclosure concern disappears, and one can achieve an 
acceptable cosmology. This has already been shown in 
a model with one extra dimension \cite{sp}, and we will 
demonstrate that the situation is the same in our model (though a detailed study 
shall form a separate report \cite{rhbsp}). 

However, it is important to stress that these massless modes which are so important for
SGC to be viable appear naturally in our framework rather than as an 
ad hoc input. We now turn to the discussion of the model. We have left the total number of 
spactime dimensions $D$ and the number of compact dimensions $p$ arbitrary, 
with different choices corresponding to various corners of the string theory moduli space. 
Later we will of course be studying the case where $D-p = d = 3$ 
(three non-compact spatial dimensions).   

\section{The Model}

We will be doing Einstein gravity in the presence of string sources. This is the late time 
scenario that has to result in SGC after the dilaton has been fixed. We 
wish to determine whether it is still possible to stabilize compact extra dimension in SGC 
after dilaton stabilization. We refer to this as 
a `late time' scenario.  
We assume that the dilaton is robustly stabilized, and can hence be taken as a constant in 
what follows. We start then by assuming that the universe has the topology of 
$\mathbb R^d\times T^p$, where $d$ is the number of non-compact dimensions and $p$ is the 
number of toroidal dimensions, both of which we leave general for the moment. We make the 
metric ansatz:
\begin{equation}
\label{metric}
ds^2 = -dt^2 + \sum_{i=1}^{D} a_i^2(t)dx_i^2 \, ,
\end{equation}     
where $D$ is the total number of spatial dimensions. The resulting Einstein equations 
($G^\mu_\nu = 8\pi G_D T^\mu_\nu$) can be recast in the form:
\begin{equation}
\label{ee}
\ddot{a}_i + \dot{a}_i\Bigl(\sum_{j\neq i} \frac{\dot{a_j}}{{a_j}} \Bigr) 
= 8\pi G_D a_i\Bigr[p_i-\frac{1}{D-1}\sum_{j=1}^{D} p_j+\frac{1}{D-1}\rho \Bigl] \, .
\end{equation}

One then has to determine what the energy momentum tensor for a gas of strings 
(with a fixed set of quantum numbers) is, in order to proceed. One can obtain this directly from the Nambu-Goto action for a single string (after hydrodynamically averaging)\cite{sp}. We offer a derivation of this result in the appendix. However, it turns out that one obtains the same answer if one was to introduce the 
following `matter' Lagrangian density:
\begin{equation}
\label{ld}
S_{string} = 
-\int \sqrt{-G} \mu_{\vec{n},\vec{w}, \vec{p}, N, {\tilde N}}(t)
\epsilon_{\vec{n},\vec{w}, \vec{p}, N,{\tilde N}} \, ,
\end{equation}
where $G$ is the determinant of the full space-time metric, 
the subscripts indicate that the number density ($\mu$) and the energy of 
a given string state ($\epsilon$) depend on its particular quantum numbers. The 
notation is as follows: $\vec{n}$ describes the momentum quantum numbers along the compact 
directions, which have been organised into a  p-dimensional vector. 
Similarly, $\vec{w}$ describes the winding quantum numbers, $N$
and ${\tilde N}$ are the 
oscillator levels of the string state and $\vec{p}$ is the center of mass momentum along the 
non-compact directions, organised into a $d = D-p$ dimensional vector.   
We see that this naively corresponds to introducing the appropriate `interaction energy' 
term to our action. If one (we shall drop the subsripts momentarily) 
factors out the metric dependence of the number density
\begin{equation}
\label{num}
\mu = \frac{\mu_0}{\sqrt{g}} \, ,
\end{equation}   
where $g$ is the determinant of the spatial part of the metric tensor, one is left with:
\begin{equation}
\label{mld}
S_{string} = -\int \sqrt{-G_{00}} \mu_0(t)\epsilon \, ,
\end{equation}
which, once we consider the thermodynamics of the system, corresponds to the treatment 
given in \cite{TV}. 

The mass of a closed string in a background where p dimensions are 
compactified on a torus, described by the metric $\gamma_{ab}$, $1 \leq a,b \leq p$, is 
given by \cite{pol} ($n_a$, $w^a \epsilon ~ \mathbb Z$): 
\begin{equation}
\label{csm1}
m^2_{\vec{n},\vec{w},\vec{p},N,{\tilde N}} = \frac{1}{R^2}\gamma^{ab}n_an_b 
+ \frac{R^2}{\alpha'^2}\gamma_{ab}w^aw^b + \frac{2}{\alpha'}(N + \tilde N - 2) \, ,
\end{equation}
where this formula is subject to the level matching constraint:
\begin{equation}
\label{lmc}
\tilde N = n_aw^a + N
\end{equation}
and $R$ describes the coordinate interval for each cycle of the torus: 
$x^a = \theta^a R$, $0 \leq \theta \leq 2\pi$, which can be set to unity 
(the physical lengths of cycles being described by the metric $\gamma$). 
We can re-write the above as:
\begin{equation}
\label{csm}
m^2_{\vec{n},\vec{w},\vec{p},N} = (n,\gamma^{-1}n) 
+ \frac{1}{\alpha'^2}(w,\gamma w) +\frac{2}{\alpha'}(N + \tilde N - 2) \, ,
\end{equation}
where $(n,w) = n_aw^a$ is the standard p-dimensional real scalar product. 
We now use (\ref{lmc}) to solve for $\tilde N$ in (\ref{csm}), which gives us the 
following expression for the energy of a closed string in such a background:
\begin{equation}
\label{cse}
\epsilon_{\vec{n},\vec{w},\vec{p},N} 
= \frac{1}{\sqrt{\alpha'}}\sqrt{\alpha' p^2_{n.c.} 
+ (n,\bar{\gamma}^{-1}n) + (w,\bar{\gamma}w) +2(n,w) + 4(N-1)} \, ,
\end{equation}
where $\bar{\gamma}_{ab} = \gamma_{ab}/\alpha'$, and $p^2_{n.c.}$ is the center of mass 
momentum along the non-compact directions. 

Notice that we are not ignoring the oscillator 
modes, as is customary in the SGC literature. This allows us to utilise the level matching 
constraint in such a way that the above expression for the mass of a string mode contains a 
perfect square when all compact dimensions are at the self dual radius 
($\bar{\gamma}_{ab} = \delta_{ab}$):
\begin{equation}
\label{sdm}
m^2 = \frac{1}{\alpha'}\Bigl[(n+w,n+w) + 4(N-1)\Bigr] \, .
\end{equation}    
This should be contrasted to the expression obtained when oscillator modes are 
neglected at the outset:
\begin{equation}
\label{crap}
m^2 = \frac{1}{\alpha'}\Bigl[ (n,n) + (w,w) \Bigr] 
\end{equation}
which only vanishes for completely unwound strings, and hence fails to capture the wound 
states that become massless at enhanced symmetry points. These states prove to have very 
special properties and should not be ignored in any study of string gas cosmology. 

Returning 
to the question of deriving the energy momentum of a string gas, one can insert the 
expression (\ref{cse}) into (\ref{ld}) to arrive at:
\begin{eqnarray}
\label{tt}
\rho_{\vec n, \vec w} &=& \frac{\mu_{0, \vec n, \vec w}}{\epsilon_{\vec n, \vec w}\sqrt{-G}} \epsilon^2_{\vec n, \vec w}\\
\label{ii}
p^i_{\vec n, \vec w} &=& \frac{\mu_{0, \vec n, \vec w}}{\epsilon_{\vec n, \vec w}\sqrt{-G}}p^2_{n.c.}/d\\
\label{aa}
p^a_{\vec n, \vec w} &=& \frac{\mu_{0, \vec n, \vec w}}{\epsilon_{\vec n, \vec w}\sqrt{-G}\alpha'}\Bigl( \frac{n_a^2}{\tilde{b}_a^2} - w_a^2\tilde{b}_a^2 \Bigr)
\end{eqnarray}    
where $\tilde b_a = b_a/\sqrt{\alpha'}$ (where $b_a^2$ is the $a^{th}$ diagonal component of the metric of the torus). Henceforth $\vec n, \vec w$ shall serve as an abbreviation for the full set of quantum 
numbers, which include $N$ and $p^2$ as well. Note that the above can be 
obtained directly from the Nambu-Goto action \cite{sp}, as shown in the appendix. We can immediately infer several qualitative conclusions from the above form of the energy momentum tensor. We see that according to (\ref{aa}), winding quantum numbers contribute a negative 
pressure along the compact directions whereas momentum quantum numbers do the opposite. 
Along the non-compact directions, the fluid of strings exerts a positive pressure as is 
typical for a gas of point particles. 

The equations of motion (\ref{ee}) for both the 
compact and the non-compact dimensions which result from the above are given by:
\begin{eqnarray}
\label{c}
\ddot{\tilde{b}}_a + \dot{\tilde{b}}_a(\sum_j \frac{\dot{a_j}}{a_j} 
+ \sum_{b \neq a}\frac{\dot{b}_c}{b_c}) 
&=& \frac{8\pi G_D \mu_{0, \vec n, \vec w}}{\alpha'^{3/2}\sqrt{\hat{G}_a}
\epsilon_{\vec n, \vec w}}\Bigl[ \frac{n^2_a}{\tilde{b}^2_a} - w_a^2\tilde{b}_a^2 
+  \frac{2}{(D-1)}[(w,\bar{\gamma}w) + (n,w) + 2(N-1)] \Bigr]\\
\label{nc}
\ddot{a}_i + \dot{a}_i(\sum_{j\neq i} \frac{\dot{a_j}}{a_j} + \sum_{b}\frac{\dot{b}_c}{b_c}) 
&=& \frac{8\pi G_D \mu_{0, \vec n, \vec w}}{\sqrt{\hat{G}_i}\epsilon_{\vec n, \vec w}}
\Bigl[ \frac{p^2_{n.c.}}{d} + \frac{2}{\alpha'(D-1)}[(w,\bar{\gamma}w) + (n,w) + 2(N-1)] 
\Bigr] \, ,
\end{eqnarray} 
where $\hat{G}_\mu$ is the determinant of the metric without the $\mu$'th diagonal element.
In the above, the indices $i = 1, ..., d$ run over the non-compact dimensions and
$a = 1, ..., p$ run over the compact ones. 
One thing to note from (\ref{c}) is that unwound strings with $n_a = w_a=0$ at 
the oscillator level $N=1$ do not contribute to the driving term (the right hand side) 
of the equation of motion for the compact dimensions. These states correspond to gravitons, 
but in general from the form of (\ref{ee}), we see that any matter which is pressureless 
along the compact dimensions (such as ordinary matter) and satisfies the equation of 
state ($p = \rho/d$) also does not contribute to the driving term for the compact 
dimensions. 

The above equations are somewhat 
artificial in that they apply to a universe filled with a 
string gas with a fixed set of quantum numbers. In general, one will have a string gas 
that consists of a superposition of strings with many different quantum numbers. In 
that case, the driving terms in the above equations of motion will become:
\begin{eqnarray}
\label{dc}
\ddot{\tilde{b}}_a + \dot{\tilde{b}}_a(\sum_j \frac{\dot{a_j}}{a_j} 
+ \sum_{b \neq a}\frac{\dot{b}_c}{b_c}) 
&=& \sum_{\vec n, \vec w} \frac{8\pi G_D \mu_{0, \vec n, \vec w}}{\alpha'^{3/2}
\sqrt{\hat{G}_a}\epsilon_{\vec n, \vec w}}\Bigl[ \frac{n^2_a}{\tilde{b}^2_a} 
- w_a^2\tilde{b}_a^2 +  \frac{2}{(D-1)}[(w,\bar{\gamma}w) + (n,w) + 2(N-1)] \Bigr]\\
\label{dnc}
\ddot{a}_i + \dot{a}_i(\sum_{j\neq i} \frac{\dot{a_j}}{a_j} + \sum_{b}\frac{\dot{b}_c}{b_c}) 
&=& \sum_{\vec n, \vec w}\frac{8\pi G_D \mu_{0, \vec n, \vec w}}{\sqrt{\hat{G}_i}
\epsilon_{\vec n, \vec w}}\Bigl[ \frac{p^2_{n.c.}}{d} 
+ \frac{2}{\alpha'(D-1)}[(w,\bar{\gamma}w) + (n,w) + 2(N-1)] \Bigr] \, .
\end{eqnarray}
The number densities $\mu_{0, \vec n, \vec w}$ are determined by physical considerations.
In the spirit of SGC and
following the treatment given in \cite{sp}, we assume  
thermal equilibrium at the initial time (when the number densities are determined).
In this case, we have
\begin{equation}
\label{numte}
\mu_{0,\vec n,\vec w} 
= \mu_{0, ref}e^{-\beta \epsilon_{\vec n, \vec w}}e^{\beta \epsilon_{ref}} \, ,
\end{equation}
where the subscript `ref' refers to some reference state which we are free to pick at our 
convenience. One has to wonder what constitutes the thermal bath that is a pre-requisite 
for any  thermodynamical treatment. We have already seen that gravitons and photons 
(matter which only exists in the non-compact dimensions) do not contribute to the driving 
term for the equations of motion for the $b_a$'s, 
and are an ideal candidate for such a thermal bath. The 
tree level processes $h_{\mu\nu} \to w + \bar w$ provide the appropriate interactions 
that thermally couple our stringy matter to this bath. With this in mind, 
(\ref{numte}) implies that the driving term for the non compact dimensions becomes:
\begin{equation}
\label{therm}
\frac{8\pi G_D\mu_{ref}e^{\beta\epsilon_{ref}}}{\alpha'\sqrt{\hat{G}_a}}
\sum_{\vec n,\vec w,N,p^2} 
\frac{e^{-\beta\tilde{\epsilon}_{\vec n,\vec w,N,p^2}/\sqrt{\alpha'}}}
{\tilde{\epsilon}_{\vec n,\vec w,N,p^2}}
\Bigl[\frac{n^2_a}{\tilde{b}^2_a} - w_a^2\tilde{b}_a^2 +  
\frac{2}{(D-1)}[(w,\bar{\gamma}w) + (n,w) + 2(N-1)] \Bigr]
\end{equation}  
Where $\tilde{\epsilon}$ is a dimensionless quantity (energy expressed in string 
units: $\epsilon = \tilde{\epsilon}/\sqrt{\alpha'}$), which according to (\ref{cse}) is a 
number of order unity for the lowest lying string states. 

With this realization of thermal equilibrium, were we to define the partition function as
\begin{equation}
\label{z}
Z(\beta,a_i,b_a) := \sum_{\vec n,\vec w,N,p^2} 
e^{-\beta \epsilon(a_i,b_a)_{\vec n,\vec w,N,p^2}} \, ,
\end{equation} 
then one could derive the components of the energy momentum tensor as follows 
(c.f. \cite{TV}):
\begin{eqnarray}
\label{pi}
P_i &=& \frac{1}{\beta}a_i\frac{\partial Z}{\partial a_i} ~~=~~ 
\frac{1}{\beta}\frac{\partial Z}{\partial \lambda_i} ~~;~~ a_i = e^{\lambda_i}\\
\label{pa}
P_a &=& \frac{1}{\beta}b_a\frac{\partial Z}{\partial b_a} ~~=~~ 
\frac{1}{\beta}\frac{\partial Z}{\partial \lambda_a} ~~;~~ b_a = e^{\lambda_a}\\
\label{r0}
E &=& \frac{-1}{\beta}{}\frac{\partial Z}{\partial \beta} \, ,
\end{eqnarray}
where $P_\mu$ (and in a similar way, $E$) is defined through the equation:
\begin{equation}
\frac{\mu_{ref}e^{\beta \epsilon_{ref}}}{\sqrt{G}} P_\mu = p_\mu 
\end{equation}

We can use this to rewrite the condition that the driving term for equations of motion 
vanishes for any particular direction, as the action of a linear differential operator on 
$Z$:
\begin{equation}
\label{int}
\Bigl[a_\mu \frac{\partial }{\partial a_\mu} 
- \frac{1}{D-1}\Bigl(\sum_\nu a_\nu\frac{\partial }{\partial a_\nu} + \frac{\partial }{\partial \beta} \Bigr)\Bigr]Z = 0 \, .
\end{equation}
One can then rephrase the question of whether or not stringy matter can stabilize extra 
dimensions in terms of the existence of solutions to this equation. Were we to find 
simultaneous solutions to this system of first order PDE's for all compact directions 
($\mu = a$, for a given $\beta$), we will have determined the nature and existence of 
stabilized extra dimensions. Although we will show further on that, in the particular regime 
we are interested in, we will have a much simpler means of proceeding, this method is 
certainly more general (though less tractable). It might be neccesary to resort to this 
method in high temperature regimes and in situations where the compactifications we study 
are not so straightforward. In fact, if all we are interested in is proving the existence 
of stabilized extra dimensions, then one could imagine reformulating this question in terms 
of the existence of solutions to this PDE, for which a great deal is already known. 

However, 
since we are interested in late times (i.e. post dilaton stabilization), we find after 
inspection of (\ref{therm}) that the summation is actually quite 
tractable. Taking a closer look at the Boltzmann factor that weights all terms in the 
summation:
\begin{equation}
\label{bw}
e^{-\frac{\beta}{\sqrt{\alpha'}}\tilde{\epsilon}} \, .
\end{equation}
we see that the argument of the exponential is the factor $\beta/\sqrt{\alpha'}$ 
multiplied by a term that is of order unity. Now for all string theories there exists a 
limiting temperature, known as the Hagedorn temperature \cite{hag}, which, independent of 
the particular theory, is always of the order $\sqrt{\alpha'}$ \cite{mb}:
\begin{equation}
\label{bh}
\beta_H \sim \sqrt{\alpha'} \, .
\end{equation}
If we set the string scale to be equal to the Planck scale, then we see that if the 
temperature is even slightly (let us say a factor of 10) below this scale, 
that is if:
\begin{equation}
\label{lt}
\frac{\beta}{\sqrt{\alpha'}} \gtrsim  \, 10 \, ,
\end{equation}    
then any term in the summation (\ref{therm}) which corresponds to an $\tilde\epsilon$ 
which is anything other than zero will contribute vanishingly. Hence, the Boltzmann weight 
approaches a window function which projects out all but the 
massless modes in the summation. That is:
\begin{equation}
\label{summa}
\sum_{\vec n,\vec w,N,p^2} \frac{e^{-\beta\tilde{\epsilon}_{\vec n,\vec w,N,p^2}/\sqrt{\alpha'}}}{\tilde{\epsilon}_{\vec n,\vec w,N,p^2}}\bigl[...\bigr] 
~~\to~~ \lambda \sum_{m^2=0}\bigl[...\bigr]
\end{equation} 
and hence the sum becomes very easy to calculate. Note that the condition (\ref{lt}) 
gives us an operational definition of 
`late times'. 

Turning to the question of evaluating the summation (\ref{therm}) in light of (\ref{summa}), 
we first need to know which quantum numbers correspond to massless states for a given form 
of the metric $\tilde\gamma$ of the torus. To do this, suppose we start out with all 
toroidal dimensions compactified at the self-dual radius ($\tilde\gamma = I$). Then the 
formula for the mass of a closed string in this background is given by (\ref{sdm}):
\begin{equation}  
\nonumber
m^2 = \frac{1}{\alpha'}\Bigl[(n+w,n+w) + 4(N-1)\Bigr] \, ,
\end{equation}
from which we read off that the massless states are those which simultaneously satisfy 
the following set of equations:
\begin{eqnarray}
\label{zero}
(n+w,n+w) &=& 4(1-N)\\
\label{lc}
N + (n,w) &\geq& 0
\end{eqnarray}
where the last equation is the level matching constraint. We immediately see that the 
only possibilities for massless states at this radius (and as it turns out all others) are 
those with oscillator levels 0 or 1. We catalogue the quantum numbers of all of these 
massless states below:

\begin{center}
\begin{tabular}{c|ccc|}
$N~$&$(n,n)~$&$(w,w)~$&$(n,w)~$\\
\hline
1&0&0&0\\
1&1&1&-1\\ 
0&1&1&1\\
0&2&2&0\\
\hline
0&1&3&0\\
0&3&1&0\\
0&4&0&0\\
0&0&4&0\\
\end{tabular}
\end{center}

It should be noted that the exact realisation of these massless modes depends on the number 
of extra dimensions available (e.g. for 1 or 2 extra dimensions, it is not possible to 
satisfy $(w,w)=3$ etc.). Of most interest to us (for reasons to be made clear shortly) are 
the first four possibilities mentioned in the above table. The first possibility 
($N=1,n=w=0$) corresponds to unwound gravitons. These states do not contribute to the 
driving term for the compact directions (\ref{therm}) at all. The second possibility 
($N=1$, $n=\pm w$, $(w,w)=1$) corresponds to singly wound strings with an equal and 
opposite quantum of momentum along the same dimension, at oscillator level 1. The third 
possibility ($N=0$, $n=w$, $(w,w)=1$) corresponds to a singly wound state at oscillator 
level zero, with one quantum of momentum of the same sign along the same dimension. The second class of states are all at oscillator level zero, and correspond to various 
multiply wound/unwound strings with/without motion along various cycles of our torus. 
It turns out that if we consider the masses of states with the quantum numbers 
tabulated above as the metric fluctuates around the self dual radius, we find that only 
the first class of states (the first four in the above table) remain massless to first 
order. We see this by perturbing the metric as follows:
\begin{eqnarray*}
\label{gp}
&\tilde{\gamma}& = I - \Delta\\
&\tilde{\gamma}^{-1}& = I + \Delta + \sum_{k=2} \Delta^k ~~~~~~~~~~~~~||\Delta|| < 1
\end{eqnarray*} 
where $\Delta$ is not neccesarily small, but has a matrix norm of less than 1 
(so that $\tilde{\gamma}^{-1}$ can be written thus). Expanding the formula for the mass 
of a closed string, we get
\begin{eqnarray}
\alpha' m^2 &=& (n,\tilde{\gamma}^{-1}n) + (w,\tilde{\gamma}w) + 2(n,w) + 4(N-1)\\ 
&=& (n+w,n+w) + 4(N-1) - (w,\Delta w) + (n,\Delta n) + \sum_{k=2}(n,\Delta^kn) \, ,
\end{eqnarray}
and hence, for the mass difference compared to the value for $\Delta = 0$,
\begin{equation}
\alpha' \delta m^2 = (n,\Delta n) - (w,\Delta w) + \sum_{k=2}(n,\Delta^kn) \, .
\end{equation}
We see that only the first class of states remain massless to first order. For the first three sets of states, where the quantum number vectors $n,w$ are equal to each other up to a sign, this is easy to see. For the states with quantum numbers given by $N=0$, $(n,n) = (w,w) = 2$, $(n,w) = 0$, only the states with entries in the same rows have vanishing fluctautions. This is because the condition $(n,w)=0$, in conjunction with $(n,n)=(w,w)=2$ implies that these states either have no, or both entries in common. The contribution from the metric fluctations would be non-vanishing in the former case, but for the latter, states of the form below remain massless to first order:

\begin{eqnarray*}
w &=& (1,1)~ n = \pm(1,-1),\\
w &=& (1,-1)~ n = \pm(1,1),\\
w &=& (-1,1)~ n = \pm(1,1),\\
w &=& (-1,-1) ~ n = \pm(1,-1),
\end{eqnarray*}

\noindent where we have only indicated the non-vanishing entries in the above. At late times, when 
$\beta/\sqrt{\alpha'} \gg 1$, all states which do not remain exactly massless close to the 
self-dual radius are projected out of the summation (\ref{therm}). To emphasize this point, 
let us consider an epoch when the temperature is two orders of magnitude below the Planck 
energy. Then the Boltzmann factor goes as $e^{-100}$ for all states that are not exactly 
massless close to the self dual radius. We will show later that at exactly the self dual 
radius, the summation over all the remaining states tabulated above sums to zero, and 
hence their effects on the dynamics of the extra dimensions is vanishing.
      
Now that we have determined the properties of 
the string states which will enter the summation (\ref{therm}), we now have to 
compute their contribution to the driving term and sum over all the possiblities. 
Recalling that the unwound graviton states ($N=1$,$n=w=0$) do not contribute to the 
driving term, we find that the contribution of the states with quantum numbers 
$N=1$, $n=\pm w$, $(w,w)=1$ to (\ref{therm}) for the $a^{th}$ compact direction is:
\begin{equation}
\label{mt}
\frac{8\pi G_D \mu_{0 ref}}{\alpha'^{3/2}\sqrt{{\hat G}_a}}\frac{2}{|p_{n.c.}|}
\Bigl[ \frac{1}{\tilde{b}_a^2} - \tilde{b}_a^2 + \frac{2}{D-1}(\sum_{c=1}^{p} \tilde{b}_c^2 
- p) \Bigr] \, ,
\end{equation}     
where if we take the reference energy to be precisely one of the massless states, 
the exponential prefactor in (\ref{therm}) cancels. The factor $2/|p_{n.c.}|$ comes from the overall degeneracy of the states which can appear either as, for example 
$w = (0,0,0...1,0...)$, $n = (0,0,0...-1,0...)$ or with the opposite signs, with the
factor $|p_{n.c.}|$ coming from the factor of energy in the denominator in the summand in (\ref{therm}). The states with quantum numbers $N=0$, $n=w$, $(w,w)=1$ sum to yield an identical contribution to the driving term. It is straightforward to show that the states within the class $N=0$, $(n,n)=(w,w)=2$, $(n,w)=0$, that remain massless to first order also yield a similar driving term, but now with the prefactor $8(p-1)$, instead of $2$. The factor $p-1$ has the combinatorial origin of being the number of ways one can pick two entries to be identical out of $p$ choices, and $8$ corresponds to the overall degeneracy of these states (as indicated above). One might be worried that introducing such states with $(w,w)=2$ might force us to consider off-diagonal elements for our toroidal metric $\gamma_{ab}$, as we now have strings that diagonally wrap the torus. It is in fact true that for a {\it single} diagonally wound string, the stress energy tensor will have off-diagonal components and that these should be matched by off diagonal elements in our toroidal metric. However, recall that we are considering a string gas, which at the point in moduli space we begin in ($\gamma_{ab} = \delta_{ab}$), will consist of massless quantum numbers which will democratically wrap along all cycles of the torus with equal probablity. That is, singly wound strings will wind along all cycles of the torus with equal probability. In addition, strings wound around more than one cycle (such as the states with $(w,w) = 2$) will also wrap any given pair of cycles with equal probability. When we sum over all the quantum numbers, we will invariably encounter summing over states with winding number vectors that are opposite in sign, which results in the cancellation of off-diagonal entries in the net stress energy tensor (see appendix). \footnote{However the general issue of off-diagonal elements of the toroidal metric (which correspond to complex structure moduli), is an important one, as one would also have to address how these are stabilized in this framework. It turns out that this same string gas that we have introduced aslo stabilizes the shape moduli of the torus. In \cite{Edna} and \cite{spd}, it was shown that the shape moduli decouple from the radial moduli, and are stabilized by the effects of the string gas in an analogous way to the radial moduli. A similar result was uncovered in \cite{alikaya}.}.   

\par

Denoting $\sum_{c=1}^{p} \tilde{b}_c^2$ as $(\tilde{b},\tilde{b})$, we then find that the total contribution to the driving term from states that remain exactly massless near the self dual radius is: 
\begin{equation}
\label{mt2}
\frac{8\pi G_D \mu_{0 ref}}{\alpha'^{3/2}\sqrt{{\hat G}_a}}\frac{(8p-4)}{|p_{n.c.}|}
\Bigl[ \frac{1}{\tilde{b}_a^2} - \tilde{b}_a^2 + \frac{2}{D-1}[(\tilde{b},\tilde{b}) - p] 
\Bigr] \, .
\end{equation}

The condition for this driving term to vanish yields the 
solution which corresponds to a stabilization of the extra dimensions. This can
also be interpreted as a global minimum of the potential for the radion with vanishing
amplitude (thus avoiding the ``no-go theorem for radion stabilization by Giddings
\cite{Giddings} which studies local minima of the radion potential with positive value). 
The condition that the driving term vanish is:
\begin{equation}
\label{dtv}
\frac{1}{\tilde{b}_a^2} - \tilde{b}_a^2 + \frac{2}{D-1}[(\tilde{b},\tilde{b}) - p] = 0 \, .
\end{equation}
We see that this can be recast as a quadratic equation for $\tilde{b}_a^2$, where the 
coefficients are formally the same for all $a$. Hence, if a solution exists, it must be the 
same for all $a$. Thus we find that the extra dimensions are isotropically stabilized if 
they are stabilized at all. We now insert the ansatz appropriate to this observation 
($\tilde{b}_a = \lambda, \forall a$) into the above to obtain the condition:
\begin{equation}
\label{dtv2}
\frac{1}{\lambda^2} - \lambda^2 + \frac{2p}{D-1}[\lambda^2 - 1] = 0 \, .
\end{equation}
Using the usual technique to solve a quadractic equation, we find a solution to be:
\begin{equation}
\label{lsol}
\lambda^2 = \frac{\frac{2p}{D-1} + \sqrt{(\frac{2p}{D-1})^2 - 4[\frac{2p}{D-1} -1] } }{2[\frac{2p}{D-1} -1]}
\, .
\end{equation}
Were we to define $c = 2p/(D-1)$, we find that the above simplifies into the formula:
\begin{equation}
\label{solv}
\lambda^2 = \frac{c + (c-2)}{2(c-1)} = 1 \, .
\end{equation}
Thus, we have shown that the extra dimensions are stabilized isotropically at the
self-dual radius.

To complete the analysis we need to show that the fixed point determined by
(\ref{solv}) is indeed a stable equilibrium point. A quick way to see this is to
observe that the potential which determines the radion dynamics is minimized at this
point. This can also be seen by inserting the driving term (\ref{mt2}) into (\ref{dc}), 
and expanding $\tilde{b}_a$ as $\tilde{b}_a = 1 + \Gamma_a$. Our equations of motion become:
\begin{equation}
\label{geom}
\ddot{\Gamma}_a + \dot{\Gamma}_a(dH + \sum_{c\neq a} \dot\Gamma_c) 
+ \frac{8\pi G_D 4(8p-4)\mu_0^{ref}}{a^d \alpha'^{\frac{3+(p-1)}{2}}|p_{n.c.}|(D-1)}[(D-2)\Gamma_a 
- \sum_{c\neq a} \Gamma_c] = 0 \, ,
\end{equation}   
where $H$ is the Hubble factor in the $d$ non-compact directions. Interpreting the driving 
term as $\partial_{a} V$ where the derivatives are taken with respect to $\Gamma_a$, we 
find that the Hessian matrix ($H_{ab} = \partial_a\partial_b V$), up to a (positive) 
factor, is given by:
\begin{equation}
\label{hess}
\begin{pmatrix}
D-2&-1&-1&\ldots&-1\\
-1&D-2&-1&\ldots&-1\\
-1&-1&D-2&\ldots&-1\\
\vdots&\vdots&\vdots&\ddots&\vdots\\
-1&-1&-1&\ldots&D-2
\end{pmatrix}
\end{equation}
where we remind the reader that this is a p-dimensional matrix. The eigenvalues of this 
matrix are $D-1-p$ ($= d-1$), and $D-1$, the former appearing once and the latter with a 
degeneracy of $p-1$. These are all clearly positive, and hence we conclude that the 
fluctuations around the self-dual radius in all directions are indeed stable.
 
Let us summarize the method and discuss the result. We determined the energy momentum tensor 
of a gas of closed strings and observed that only the massless string modes will be present 
in any significant number if we start the evolution in a hot thermal state (an observation 
that will also be crucial for the success of the late time phenomenology). We find that, 
after accounting for all the relevant states, their quantum numbers and their degeneracies, 
we end up with a driving term (\ref{mt2}) for the radion fields which stabilizes any number 
of extra dimensions at the self-dual radius. This result is a non-trivial result 
since it was obtained in the setting of general relativity, i.e. with fixed dilaton. Thus,
we have shown that SGC provides a mechanism for stabilizing extra dimensions in the present 
epoch. 

\par
It is a natural question to ask whether or not it was neccesary to introduce stringy states in order to affect this stabilization, as one could imagine any such states which become massless at special points of moduli space could have the same effect. This question will certainly occupy us in future work. However at present, we wish to point out that (as we will see further on), it is a property peculiar to extended objects that we can obtain moduli stabilization in such a way as to be consistent with observational bounds from late times.  
\par

We wish to end this section with a comment on the states which are massless
at the self-dual radius but whose energy is not minimized at this point
(the second class of quantum numbers in the previous table). Since they are tachyonic
either for small or for large radii, they should probably be excluded from consideration
from the outset. Even if they are included, however, we note that their effects cancel in 
the driving term of the compact dimensions at the self-dual radius. This can easily be seen 
by realising that these states appear in T-dual combinations 
($\vec w \leftrightarrow  \vec n$). Considering the driving term at the self-dual radius:
\begin{equation}
\label{thermal}
\frac{8\pi G_D\mu_{ref}}{\alpha'^{3/2}\sqrt{\hat{G}_a}|p_{n.c.}|}
\Bigl[n^2_a - w_a^2 + \frac{2}{(D-1)}[(w,w) + (n,w) + 2(N-1)] \Bigr]
\end{equation}  
we see that T-duality will ensure that the first two terms cancel each other in 
summing over all states. It is easy to check, after accounting for the correct degeneracies 
of each set of quantum numbers, that the factors grouped in the inner square brackets also 
sum to zero. This completes the demonstration of stabilization at the self dual radius. 

\section{Spacetime Dynamics of the Non-Compact Dimensions}

We would now like to consider the resulting cosmology for the non-compact dimensions after 
the extra dimensions have been stabilized (in this section, we will take $d=3$ independent 
of the choice for $D$ and $p$). Before we do this we would like to discuss several 
outstanding issues that should not be overlooked. Our first issue concerns an important 
consistency check concerning general relativity: we should check that our energy momenutum 
tensor is consistent with the covariant conservation of the Einstein tensor:
\begin{equation}
\label{cov}
\nabla_\mu T^\mu_\nu = 0 \, .
\end{equation} 
This condition yields the following series of equations (one for each value of the 
index $\nu$ in the above):
\begin{eqnarray}
\label{cont}
\dot \rho + \sum_{i=1}^{d} \frac{\dot{a}_i}{a_i}(\rho + p^i) 
&+& \sum_{a=1}^{p} \frac{\dot{b}_a}{b_a}(\rho + p^a) = 0\\
\partial_i p^i &=& 0\\
\partial_a p^a &=& 0 \, .
\end{eqnarray}
The first condition is none other than the continuity equation, and is trivially satisfied 
by (\ref{tt})-(\ref{aa}). This arises from the time derivative of $\rho$ precisely 
cancelling the terms proportional to the Hubble factors. The remaining equations are also 
trivially satisfied as a consequence of the spatial homogeneity of our setup. 

The second issue concerns the equation of state parameter for the pressure along the 
non-compact directions. We know from (\ref{ii}) that the pressure along the non-compact 
directions of this string gas is always positive. However, we see from (\ref{aa})
that the pressure along the compact directions can be either negative or 
positive. If we want to avoid violations of the dominant energy condition (DEC) we must 
ensure that the equation of state parameter $\omega = p/\rho$ remains bounded from below:
\begin{equation}
\label{om}
-1 \leq \omega \, .
\end{equation}  
Consider now the relationship between (\ref{tt}) and (\ref{aa}) for the states that we have found give us stabilization (those with quantum numbers $\vec n = \pm \vec w, (w,w)=1, N=1$ or $\vec n = \vec w, (w,w)=1, N=0$). If $p^a = \omega^a\rho$, then
\begin{equation}
\label{om2}
\omega^a = \frac{\tilde{b}_a^{-2} - \tilde{b}_a^2}{\tilde{b}_a^{-2} 
+ \tilde{b}_a^2 - 2 + \alpha'p^2_{n.c.}} \, .
\end{equation} 
Since the string states we are considering are massless when the scale factor $\tilde{b}_a$ 
is at its self-dual value, they will have non-zero momentum along the non-compact 
directions, which will assume its thermal expectation value if we are in thermal 
equilibrium. If we are in a sufficiently hot regime, we can always ensure that $\omega^a$ 
satisfies
\begin{equation}
\label{om3}
-1 \leq \omega \leq 1 \, ,
\end{equation}  
where, long after the stabilization has been achieved, and the ambient temperature has 
cooled down considerably, one has a robust stabilization mechanism that keeps the compact 
dimensions locked at the self dual radius, where the equation of state parameter vanishes. 
Hence we can easily arrange a situation where the DEC is not violated in our model even 
though the compact dimensions are undergoing damped bounces. This is a novel result in the 
context of GR and was uncovered first in our study of this model in the case where we only 
had one extra dimension \cite{sp}.

Turning now to the issue of the resulting spactime dynamics of the non-compact directions, 
which we take to be homogenous and isotropic, we remind the reader of the Einstein 
equations applied to our anisotropic metric, Eq. (\ref{ee}).
We consider a situation where the dominant matter component of the universe as a whole is 
in the form of radiation; that is, matter which has no pressure along the compact 
directions and which satisfies the equation of state
\begin{equation}
\label{rad}
p = \rho/3 \, .
\end{equation} 
Considering the effect of this matter on the dynamics of the compact dimensions, we have
\begin{equation}
\label{eec}
\ddot{b}_a + \dot{b}_a\Bigl(3H + \sum_{c\neq a}^{p} \frac{\dot{b}_c}{b_c} \Bigr) 
= 8\pi G_D b_a\Bigr[ -\frac{1}{D-1}dp+\frac{1}{D-1}\rho \Bigl] \, .
\end{equation}
Thus, we see that such matter does not contribute to the dynamics of the compact dimensions.
Hence, the stringy matter studied in the previous section will be the only factor at play 
in the dynamics of these dimensions, except of course for the Hubble damping factor due
to the expansion of the large dimensions. To put it in another way, it is consistent with 
the stabilization mechanism we have studied to have radiation like matter drive the 
expansion of the non-compact dimensions. We see through (\ref{00}), (\ref{ee}) and 
(\ref{cont}), that the standard FRW expansion of the non-compact dimensions results:
\begin{eqnarray}
\rho &=& \frac{3}{8\pi G_D}H^2\\
\frac{\ddot{a}}{a} &+& 2H^2 = 8\pi G_D p\\
\dot \rho &+& 3H(\rho + p) = 0 \, .
\end{eqnarray}
Hence, it is easy to realise a post-stabilization radiation dominated 
FRW expansion of the universe. We can also obtain dust dominated evolution in our model, but this is not such a straightforward matter. In fact, obtaining a dust driven expansion results in a definite prediction of SGC: if string gases are indeed responsible for 
present day stabilization of extra dimensions in our model, then the dark matter will neccesarily have to be extra dimensional in nature. This issue was studied in \cite{sp} and we repeat the argument here. Considering (\ref{ee}) for a compact dimension, any matter which only exists in the non-compact dimensions ($p^a = 0$) which satisfies the equation of state $p^i = 0$ (i.e. all pressures vanish) will neccesarily lead to expansion of the compact directions and will derail any stabilization mechanism we might have had in 
place. Hence, if we take the dominant matter content of the universe to have the equation of state of dust, then it must neccesarily exert pressure along the compact directions which satisfies the following equation of state when the compact dimensions are all at the 
self-dual radius:
\begin{equation}
\label{dm1}
\Bigl[p^a - \frac{1}{D-1}\sum_{b=1}^p p^b + \frac{1}{D-1}\rho\Bigr]_{b_a = \sqrt{\alpha'}} 
= 0 \, .
\end{equation}
This implies, since we have isotropically stablized at the self-dual radius, that this matter satisfies the equation of state 
\begin{equation}
\label{dm2}
r = -\rho/(d-1) = -\rho/2 \, ,
\end{equation} 
where $r$ is the pressure along any of the compact dimensions.
There is a candidate within our framework for 3 large spatial dimensions (see \cite{sp}), 
namely the stringy states with the quantum numbers
\begin{equation}
\label{dmc}
|p_{n.c.}|=0, N=2; n^a=0, w^a=\pm 2 \, .
\end{equation}
It can easily be checked that such matter satisfies the required equation of state. 
These states contribute to the stability of the extra dimensions rather feebly when 
phenomenological bounds are accounted for \cite{sp}, but preserve the stability of the 
extra dimensions nevertheless (the massless states being dominant in their contribution to 
the stabilization mechanism). Hence we can take these states as a candidate for the dark 
matter responsible for our present FRW expansion \footnote{String winding states
as candidates for dark matter were recently also considered in \cite{Gubser1,Gubser2}.}

The resulting FRW equations, once the extra dimensions have stabilized are given by:
\begin{eqnarray}
\rho &=& \frac{3}{8\pi G_D}H^2\\
\frac{\ddot{a}}{a} &+& 2H^2 = 8\pi G_D \frac{(p+2)}{2(D-1)}\rho\\
\dot \rho &+& 3H(\rho + p) = 0
\end{eqnarray}
where $p$ on the right hand ride of the above refers to the number of compact dimensions. 
Hence we have shown that any epoch of late time FRW cosmology can result from our model 
post stabilization. We now turn to a brief discussion of the phenomenology of this model.

\section{Phenomenology}

Most of what is to appear in this section appears in \cite{sp}, where the phenomenology of 
a string gas used to compactify one extra dimension at the self dual radius is discussed.  
Since the string modes used in this work are an immediate generalization of what was
used in \cite{sp} for one extra dimension to the case of many extra dimensions, the 
resulting phenomenological bounds on the scenario from the point of view of late time 
cosmology are the same. Since we plan to give a detailed study of the phenomenology of 
this model in a future work \cite{rhbsp}, we here present only the briefest discussion. 

There are three key aspects to our phenomenology that we need to discuss: the first being 
that we would like not to overclose the universe with the fluid of closed strings 
(which behaves like hot dark matter from the 4-d perspective). The second aspect, 
is that we do not want to have 
too few of these strings such that the stabilization mechanism is ineffective. In words, we 
would like to show that it is possible to introduce an effective stabilization mechanism 
without overclosing the universe. The third aspect is that we do not want the dynamics of 
the extra dimensions from the 4-d perspective to introduce any long range scalar 
interactions (no fifth forces).

As a starting point, consider (\ref{geom}), which in normal coordinates has the form:
\begin{equation}
\label{damp}
\ddot{\Gamma} + 3H\dot{\Gamma} + k\Gamma = 0 \, .
\end{equation}
The value for $k$ (the spring constant) will differ by a factor of order unity depending 
on precisely which mode we are looking at, but for an order-of-magnitude estimate this 
factor is irrelevant. An upper and a lower bound on this spring constant result (in
the case of the lower bound) from requiring that the stabilization mechanism be effective, 
and (in the case of the upper bound) from requiring that the metric varies on a time scale
that is many orders of magnitude (let us say $10^6$) longer than the string scale \cite{sp} 
(otherwise the effective field theory analysis would not be justified). 
The lower bound is given by the value for $k$ which yields critical damping , 
$k_c = 9H^2/4$ and the upper bound by the string tension. Thus, we require
\begin{equation}
9H^2/4 \leq k \leq 10^{-6}/(2\pi\alpha') \, . 
\end{equation}
{F}rom (\ref{geom}) this implies
\begin{equation}
\label{d2}
H^2 \leq \frac{8\pi G \mu_0^{ref}}{a^3 \alpha'^{\frac{3+(p-1)}{2}}|p_{n.c.}|} 
\leq \frac{10^{-6}}{2\pi\alpha'} \, ,
\end{equation} 
neglecting factors of order unity. Using the relationship
\begin{equation}
\label{nchd}
G_{D}= G_{3}\times Vol ~ T^p = G_3(2\pi\sqrt{\alpha'})^{p/2}
\end{equation}
between the higher dimensional gravitational constant $G_D$ and Newton's constant
$G_3$ in our $3+1$ dimensional space-time,
and setting the string scale to the Planck scale implies
\begin{equation}
\label{nchd2}
2\pi\alpha' = G_3 \, .
\end{equation}  
Thus, (\ref{d2}) becomes:
\begin{equation}
\label{d3}
H^2 \leq \frac{\mu_0}{a^3|p_{n.c.}|} \leq \frac{10^{-6}}{2\pi\alpha'} \, .
\end{equation}

Furthermore, from (\ref{tt}) we see that in order not to overclose the universe with 
these strings we require that the 4-dimensional energy density 
($\rho_4 = \rho_D\times Vol~ T^p$) be several orders of magnitude less than the critical 
density:
\begin{equation}
\label{over}
\rho = \frac{\mu_0|p_{n.c.}|}{a^3} \leq 10^{-4}\rho_{crit} \, .
\end{equation}
Taking $\rho_{crit} = 10^{-29}g/cm^3$, we find that this bound translates into
\begin{equation}
\label{over2}
\mu_0 \leq 10^{-4}10^{-10}eV^4|p_{n.c.}|^{-1} \, .
\end{equation}
If we parametrize the momentum along the non-compact directions as
\begin{equation}
\label{ppp}
|p_{n.c.}| = 10^{-\gamma} eV = 10^{-\gamma}eV \, ,
\end{equation} 
then (\ref{over2}) becomes:
\begin{equation}
\label{pp1}
\mu_0 \leq 10^{\gamma - 41}GeV^3 \, ,
\end{equation}
which is stronger than the upper bound in (\ref{d3}), whereas the lower bound implies:
\begin{equation}
\label{pp2}
\mu_0 \geq H^2|p_{n.c.}| \sim 10^{-93-\gamma}GeV^3  \, .
\end{equation} 
Hence, we conclude that:
\begin{equation}
\label{pp3}
10^{-93-\gamma}GeV^3 \leq \mu_0 \leq 10^{-32+\gamma}GeV^3 \, ,
\end{equation}
which is easily satisfied. The remaining constraint comes from requiring that from the 
perspective of the effective 4-d theory, the masses of the fluctuations are sufficiently 
high so as not to mediate long range `fifth forces'. Observational bounds require the 
mass of these scalars to be greater than $10^{-12}GeV$. Since the spring constant 
corresponds to the masses squared of these fluctuations, (\ref{d3}) implies that:
\begin{equation}
\label{ffc}
\mu_0 \geq 10^{-33-\gamma} GeV^3 \, ,
\end{equation} 
which, in conjunction with (\ref{pp3}), leads to the condition 
\begin{equation}
\label{final}
10^{-33-\gamma}GeV^3 \leq \mu_0 \leq 10^{-41+\gamma}GeV^3 \, .
\end{equation}
This is easy to satisy for any $\gamma \geq 4$. 

Hence, we have demonstrated that the stabilization mechanism, because of some very novel 
aspects of the string gas energy momentum tensor, not only offers a robust stabilization 
mechanism, but does so in a way that is phenomenologically consistent. That is, we can 
obtain this stabilization without overclosing the universe and violating any 
fifth force constraints. This is in marked difference with previous attempts at obtaining 
stabilizing effects for extra dimensions by introducing new fields-- where the energy 
density that appears in the energy momentum tensor and the mass of the resulting scalar 
field (or the spring constant for the radion fluctuations) in the effective field theory 
are usually proportional to each other. We see that for stringy matter, they are markedly 
different in that $\rho \propto |p_{n.c.}|$, whereas $k \propto |p_{n.c.}|^{-1}$. A more thorough study of the phenomenology of this scenario, where more issues (including a demonstration of the stability of this model to fluctuations) will be 
tackled, will the subject of a future report \cite{rhbsp}. 

\section{Conclusions}

In this report, we have studied the effects of a gas of closed strings on the dynamics of a homogeneous but anisotropic space-time described by General Relativity (fixed dilaton) where several dimensions are toroidally compactified. The modes that turn out to be relevant 
are the string modes which are massless at the self-dual radius. In the context of a hot early universe we show, based on thermodynamical considerations, that these are the dominant modes. We demonstrated that these modes lead to the isotropic stabilization of the
extra dimensions, thus generalizing the results of \cite{sp} which were derived in the case of only one extra dimension.
We also showed that the dominance of these massless modes is crucial to the late-time phenomenological viability of the scenario, and leads to a scenario  consistent with several observational bounds. We feel that this formalism is the begining of a promising avenue of research, which we believe may result in a complete and testable 
model of the universe. We already have promising indications that this framework is capable of modelling a non-singular bouncing cosmology \cite{brown}, and potentially provides a stringy mechanism for the generation of metric fluctuations. Our ultimate hope is to extract testable predictions of this formulation of string gas cosmology, the prospects for which seem very promising.

\section{Acknowledgmenets}

We wish to thank Thorsten Battefeld, Tirthabir Biswas and Scott Watson for many useful discussions. SP wishes to thank the people, and the city of Montr\'eal for all the inspiration, kindness and intellectual stimulation they have provided during the time this work was undertaken. This work is supported by funds from McGill University,
by an NSERC Discovery Grant (at McGill) and (at Brown) by the US Department of Energy under Contract DE-FG02-91ER40688, TASK A.

\appendix{}

\section{The string gas energy momentum tensor}

In this appendix, we derive the energy momentum tensor of a string gas from micro-physical considerations. That is to say, we first arrive at the space-time energy momentum tensor of a single string, after which we perform a hydrodynamical averaging to obtain the result for a string gas. We wish to comment that there are several non-trivial issues to address when considering string propagation on a time dependent background. Intuitively however, one is tempted to conclude that provided the metric of space-time is varying on time scales much longer than the string scale, we should be able to proceed (as we do here) in the `adiabatic' approximation. We shall take this for granted in what follows, refering the reader to \cite{sp} for a justification of this approximation.   

\par

To begin with, consider the Nambu-Goto action for a single string:

\eq{sng}{S = -\frac{1}{2\pi\alpha'}\int d^2\sigma \sqrt{-h},}

\noindent with the worldsheet metric $h_{ab}$ defined by:

\eq{wsm}{h_{ab} = g_{\mu\nu}(X)\partial_aX^\mu\partial_bX^\nu.}

\noindent The space-time metric, which in the above context is generically a function of the worldsheet fields $X^\mu$, is assumed to have the form:

\begin{eqnarray}
\label{goo}
g_{00} &=& g_{00}(X^0),\\
\label{gij}
g_{ij} &=& \delta_{ij}a^2(X^0),\\
\label{gab}
g_{ab} &=& \gamma_{ab}(X^0). 
\end{eqnarray}

\noindent Variation of (\ref{sng}) with respect to $g_{\mu\nu}$ gives us the {\it spacetime} energy momentum tensor through the expression:

\eq{stemt}{T^{\mu\nu} = \frac{2}{\sqrt{-g}}\frac{\delta S}{\delta g_{\mu\nu}}.}

\noindent Arbitrary variations of the background metric induce variations of the worldsheet metric in the following manner:

\eq{relvar}{\delta^{\lambda\beta}g_{\mu\nu} = \delta^{\lambda}_{\mu} \delta^{\beta}_{\nu}\delta^{D+1}(X^\tau - y^\tau)~~\to~~\delta^{\lambda\beta} h_{ab} = \partial_aX^\lambda\partial_bX^\beta\delta^{D+1}(X^\tau - y^\tau).}

\noindent The unmatched indices $\lambda$ and $\beta$ mean that we perturb only these components of the metric tensor, and $\delta^{D+1}(X^\tau - y^\tau)$ is a delta function in $D+1$ space-time dimensions. This variation results in an expression for the energy-momentum tensor for a single string:

\eq{emt1s}{T^{\mu\nu} = \frac{-1}{\sqrt{-g}2\pi\alpha'}\int d^2\sigma \sqrt{-h}h^{ab}\partial_aX^\mu\partial_bX^\nu \delta^{D+1}(X^\tau - y^\tau).}

\noindent We now pick a gauge to work in. We choose to work in conformal gauge, defined by:

\eq{cg}{h_{ab}= \lambda \begin{pmatrix} -1&0\\0&1\end{pmatrix},}

\noindent where we keep this up to an arbitrary positive factor. From (\ref{wsm}), we see that this gauge choice implies the conditions:

\eq{wcc1}{g_{\mu\nu}\dot X^\mu\dot X^\nu + g_{\mu\nu}X'^\mu X'^\nu = 0}
\eq{wcc2}{g_{\mu\nu}\dot X^\mu X'^\nu = 0.}

\noindent Even though we are in a (weakly) time-dependent background (in particular one that is not flat), it can be shown that one can still make this gauge choice simultaneous with the condition:

\eq{tdo}{X'^0 = 0,}

\noindent where the prime denotes differentiation with respect to the spacelike worldsheet co-ordinate. We will use these conditions repeatedly in what follows.
\par
Upon examining (\ref{emt1s}), we see that in order to make use of the delta functions in the integrand, one has to to use the following transformation:

\eq{jaco}{d^2\sigma = \frac{dX^0 dX^a}{|\dot X^0||X'^a|},}

\noindent where $X^a$ is the string co-ordinate field along any wound compact direction (which one we pick will turn out to be insignificant). The terms in the denominator arise in the evaluation of the Jacobian of this transformation, subject to (\ref{tdo}). Note that we picked the particular co-ordinates $X^0$ and $X^a$ because they are monotonic functions of $\sigma^0$ and $\sigma^1$ respectively. Using the constraints (\ref{wcc1}) and (\ref{wcc2}), we see that:

\eq{wcc3}{\dot X^0 = \frac{2\pi\alpha'}{\sqrt{-g_{00}}}\sqrt{g^{ij}P_iP_j + \frac{1}{(2\pi\alpha')^2}g_{ij}X'^iX'^j}~,}

\noindent where we have used the fact that in conformal gauge, we have:

\eq{mom}{P_\mu = \frac{g_{\mu\nu}\dot X^\nu}{2\pi\alpha'}.}

\noindent The expression in the sqaure root in (\ref{wcc3}) is given by the $L_0$ constraint in our constraint algebra \cite{pol} (see also (\ref{csm1}) and (\ref{lmc})), and is equal to the energy of the closed string:

\eq{energya}{\epsilon = \sqrt{|p_{d}|^2 + (n,\gamma^{-1} n) + \frac{1}{\alpha'^2}(w,\gamma w) + \frac{1}{\alpha'}[2(n,w) + 4(N-1)]},}

\noindent where the worldsheet zero modes give us the contributions $|p_d|$ for momentum along the non-compact directions, as well as the terms containing the winding and momentum quantum numbers for the compact dimensions. All the other Fourier modes give us the oscillator contributions. We can write the above as:

\eq{wcc4}{\dot X^0 = \frac{2\pi\alpha'}{\sqrt{-g_{00}}}\epsilon.} 
 
\noindent As for the second factor (\ref{jaco}) entering the Jacobian, we see that for any string wound $w$ times around the $a^{th}$ direction:

\eq{wcc5}{X'^a = w^a.}

\noindent Recall that one has to sum over all zeroes of the argument of the delta function along the $a^{th}$ direction when evaluating the integral (\ref{emt1s}), after implementing the change of variables (\ref{jaco}). Thus the contribution (\ref{wcc4}) is cancelled by the string winding $w^a$ times around the $a^{th}$ direction, as the argument of the delta function is zero precisely $w^a$ times. Note that this is how the choice of which cycle we take in evaluating the change of variable ends up being inconsequential. However, the choice is residually implicit in which of the wound co-ordinates $X^a$ remain in the $D-2$ delta functions left over after integrating over the worldsheet. This will prove to be irrelevant after we hydrodynamically average to obtain the result for a gas of strings.  Hence we evaluate (\ref{emt1s})-- using the gauge fixing conditions and the results just obtained (\ref{wcc3}) - (\ref{wcc5})-- as:
 
\begin{eqnarray}
T^0_0 &=& \frac{\epsilon}{\sqrt{g_s}}\delta^{D-1}(X^\tau - y^\tau),\\
T^i_i &=& \frac{p^ip_i}{\epsilon\sqrt{g_s}}\delta^{D-1}(X^\tau - y^\tau),\\
T^a_a &=& \frac{1}{\epsilon\sqrt{g_s}}\Bigl(\frac{n_a^2}{b_a^2} - \frac{w_a^2b_a^2}{\alpha'^2}\Bigr)\delta^{D-1}(X^\tau - y^\tau),\\
\label{tac}
T^a_c &=& \frac{1}{\epsilon\sqrt{g_s}}\Bigl(\frac{n_a n_c }{b_a b_c} - \frac{w_a w_c b_a b_c}{\alpha'^2}\Bigr)\delta^{D-1}(X^\tau - y^\tau),
\end{eqnarray}

\noindent where $g_s$ is the determinant of the spatial part of the metric, and $\epsilon$ is defined by (\ref{cse}). We now hydrodynamically average as follows: keeping the quantum numbers $p_d^2, \vec w, \vec n$ and $N$ fixed, we sum the contributions over a distribution of such strings, where the momentum along the non-compact directions is distributed isotropically. We note that according to (\ref{cse}), a wound string with quantum number vectors $\vec w$, $\vec n$ will have the same energy as a string with both vectors with the opposite sign. Moreover, for diagonally wound strings, this energy will be the same were we to keep the winding number around any given cycle fixed, whilst winding around any other cycle oppositely, provided that we also reverse the sign of the momentum quantum number corresponding to that cycle (this is so that the term $(n,w)$ remain invariant). Hence, the off diagonal terms in the above will cancel out when summing over a gas of strings. In this way, our hydrodynamical averaging results in the term (\ref{tac}) dropping out, yielding (\ref{tt})-(\ref{aa}):

\begin{eqnarray}
\label{rho3}
\rho &=& \frac{\mu_0\epsilon}{\sqrt{{g}_s}},\\
\label{pi3}
p^i &=& \frac{\mu_0}{\sqrt{{g}_s}\epsilon}|p_{d}|^2/d,\\
\label{pa3}
p^a &=& \frac{\mu_0}{\sqrt{{g}_s}\epsilon}[\frac{n_a^2}{{b}_a^2} - \frac{w_a^2 b_a^2}{\alpha'^2}].
\end{eqnarray}

\noindent As is easily checked,  we note that one would obtain this same result if one were to introduce the following action for the string gas:

\eq{sgact}{S = -\int d^{D+1}x\sqrt{-g_{00}}\mu_0\epsilon~.}

\end{document}